\documentclass{pasa}%

\title[Optical Modulation in 4U 1543$-$624]{Optical Modulation in the X-Ray Binary 4U 1543$-$624 Revisited\thanks{This paper includes data gathered with the 6.5 meter Magellan Telescopes located at Las Campanas Observatory, Chile.}}
\author[Wang et al.]{Z. Wang$^1$, A. Tziamtzis$^1$, D. L. Kaplan$^2$ \and D. Chakrabarty$^{3}$\\ 
\affil{$^1$Shanghai Astronomical Observatory, Chinese Academy of Sciences, 
80 Nandan Road, Shanghai 200030, China}%
\affil{$^2$Department of Physics, University of Wisconsin-Milwaukee,
1900 E. Kenwood Blvd., Milwaukee, WI 53211, USA}
\affil{$^3$Kavli Institute for Astrophysics and Space Research, Massachusetts Institute of Technology,
77 Massachusetts Ave., Cambridge, MA 02139, USA}}%
\jid{PASA}
\doi{10.1017/pas.\the\year.xxx}
\jyear{\the\year}

\usepackage[authoryear]{natbib}
\bibpunct{(}{)}{;}{a}{}{,}
\newcommand{\apj}{ApJ}
\newcommand{\apjl}{ApJ}

\newcommand{\aap}{A\&A}

\newcommand{\mnras}{MNRAS}

\begin{document}%
\begin{abstract}
The X-ray binary 4U 1543$-$624 has been provisionally identified as an ultracompact
system with an orbital period of $\simeq$18~min. We have carried out
time-resolved optical imaging of the binary to verify the
ultra-short orbital period. Using 140\,min of high-cadence $r'$-band photometry we
recover the previously-seen sinusoidal modulation and determine a
period $P=18.20\pm0.09$\,min.  In addition, we also see a 7.0$\times
10^{-4}$\,mag\,min$^{-1}$ linear decay, likely related to variations
in the source's accretion activity.  Assuming that the sinusoidal
modulation arises from X-ray heating of the inner face of the
companion star, we estimate a distance of 6.0--6.7\,kpc and
an inclination angle of 34$^{\circ}$--61$^{\circ}$ (90\% confidence)
for the binary. Given the stability of the modulation we can confirm that
the modulation is orbital in origin and 4U 1543$-$624 is an
ultracompact X-ray binary.
\end{abstract}
\begin{keywords}
binaries: close --- stars: individual (4U 1543$-$624) --- X-rays: binaries --- stars: low mass --- stars: neutron
\end{keywords}
\maketitle%
\section{INTRODUCTION }
\label{sec:intro}

Among the $\sim$200 known low-mass X-ray binaries (LMXBs;
\citealt{lvv07}) that consist of a compact star (either a neutron star
or a black hole) accreting via a disk from a low-mass
Roche-lobe-filling companion, there is a class called ultracompact
binaries. Unlike the majority of LMXBs, in which the companions are
ordinary, hydrogen-rich stars, the companions in ultracompact binaries
have extremely low mass and are hydrogen-poor and/or degenerate
\citep*{nrj86,ynv02}.  While there is a minimum orbital period around
80~min set by the size of a Roche-lobe-filling companion for ordinary
LMXBs (\citealt{ps81}; \citealt*{rjw82}), ultracompact binaries can
evolve to ultra-short orbital periods of few minutes
\citep*{prp02,nr03}.  The ultracompact LMXBs represent an extreme
outcome of stellar and binary evolution. They may be a significant
source of low-frequency gravitational waves (e.g.,
\citealt{nyp01,rui+10}), are likely progenitors of ``black widow''
pulsar systems (e.g., \citealt{bdh14}), and may end up like
PSR~J1719$-$1438, a pulsar with a high-density, planet-mass companion
\citep{bai+11}.

Most of the known ultracompact binaries were discovered directly from
orbital signals in their X-ray emission (e.g., \citealt{vvn12};
\citealt{car+13}; and references therein).  Indirect evidence, such as
X-ray and/or optical spectral features (\citealt*{jpc01};
\citealt{nel+04}; \citealt{wan04}) or low optical--to--X-ray flux
ratios (\citealt*{dma00}; \citealt{bas+06}; \citealt*{ijm07}), has
also led to a number of candidate ultracompact systems. For such
candidates, we still require verification of their ultracompact nature
through detection of some periodic modulation.  With that, the properties of
the binary systems can then be determined (e.g. \citealt{wc04}).  The
modulation need not be of X-rays: in an ultracompact binary, the
companion star is tidally locked such that rotation is synchronous
with the orbit.  As a result, the inner surface of the companion is
heated by strong X-ray emission from the central compact star, and as
the visible area of the heated surface varies as a function of orbital
phase, both potentially resulting in orbital modulation of the optical flux.
\begin{figure}
\centering
\includegraphics[width=8cm, angle=0]{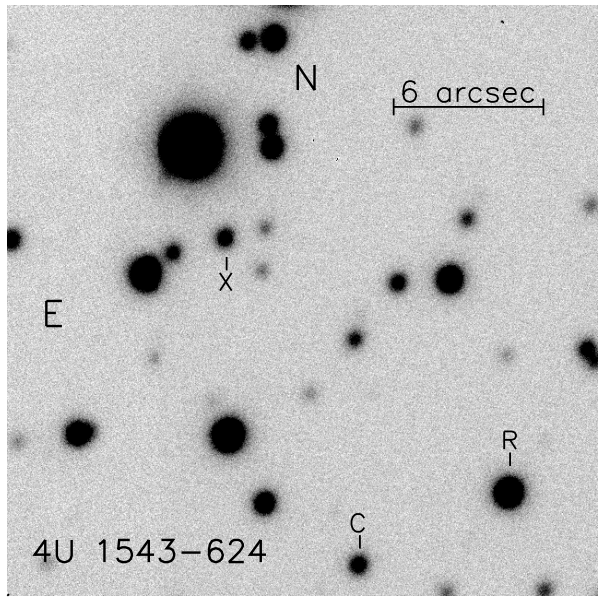}
\caption{$r'$-band image of the 4U 1543$-$624 field.
The X-ray binary target, check star, and bright reference star are marked by
$X$, $C$, and $R$, respectively.}
\label{fig:fld}
\end{figure}

We thus have conducted time-resolved imaging observations of several 
candidate systems.  The orbital periods of the LMXBs 4U~1543$-$624 and 
2S~0918$-$549 were reported previously \citep{wc04,zw11}, with both  candidates 
initially identified on the basis of their X-ray spectral 
features by \citet{jpc01}. 
In \citet{wc04}, the orbital
period of 4U~1543$-$624 was determined to likely be 18.2~min, 
revealed by the flux modulation with a fractional semi-amplitude of 8\% at 
optical $r'$ band.  However, this periodicity has not  been
confirmed through independent observations.  Given that  X-ray
binaries are often highly variable, we wished to ascertain whether the
18.2\,min period identified previously was in fact the orbital period,
or if it was some other temporary modulation of the stellar brightness.
We therefore carried out a second epoch of time-resolved 
photometry of the binary  in 2008, for the purpose of 
confirming the periodicity and examining the stability of the modulation. 
In this paper, we report the results from our observation.

\section{Observation and Data Reduction}
\label{sect:Obs}

The optical imaging observation was carried out on 2008 July 24, using
the Raymond \& Beverly Sackler Magellan Instant Camera (MagIC) on the
6.5-m Magellan/Clay telescope at Las Campanas observatory in Chile.
Unlike most MagIC observations, we used a new CCD detector for these
observations that had a very fast readout to maximize observing
efficiency.  The detector was a 1024$\times$1024\,pixel E2V CCD,
offering a field of view of 40$^{\prime \prime}$ with a pixel scale of
0.037$^{\prime \prime}$ per
pixel. A Sloan $r'$ filter was used for imaging.  In total we obtained
$121 \times 60\,$s images of the 4U 1543$-$624 field. The total time
span was approximately 140 min.  During the observation the seeing
conditions were not stable, varying from $\simeq$0.55$''$ in the
beginning to $\simeq$0.75$''$ at the end of the observation.  
\begin{figure*}
   \centering
   \includegraphics[width=12.0cm, angle=0]{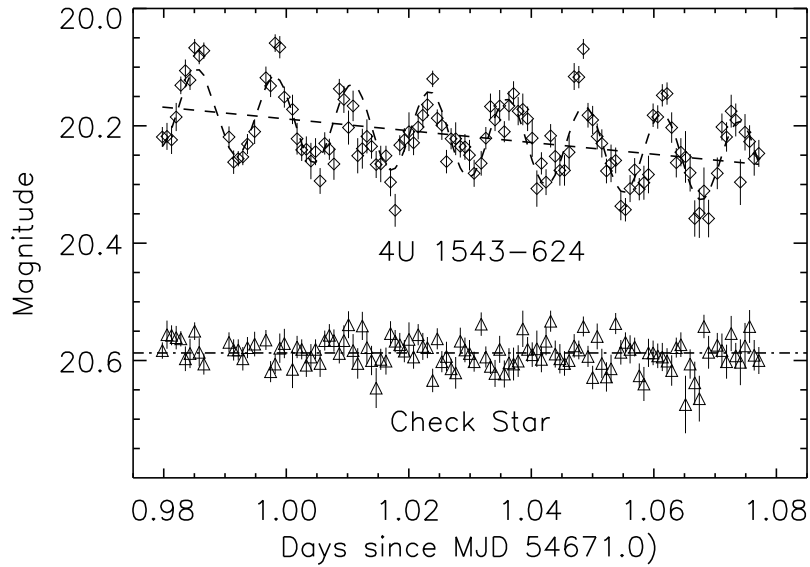}
   \caption{Optical $r'$ light curve of 4U~1543$-$624. Periodic
modulation is clearly visible. For a comparison, the light curve
of an in-field check star is also shown. The best-fit function of
a sinusoid plus a linear decay is shown as the dashed curve, and
the linear decay is indicated by the dashed line.
}
\label{fig:lc}
\end{figure*}

For the reduction of the data we used standard procedures within IRAF.
The raw images were bias subtracted and flat-field corrected.
Magnitude measurements of the target and nearby sources were obtained
using the point spread function (PSF) photometry tasks from IRAF's
DAOPHOT package.  An example image is shown in Figure~\ref{fig:fld}.
In order to eliminate any systematic variations, we used differential
photometry relative to a non-variable, bright star (star $R$ in
Figure~\ref{fig:fld}).  We also examined another non-variable star
which has a similar brightness to that of 4U~1543$-$624 to verify our
results (the same star, star $C$ in Figure~\ref{fig:fld}, as was used
in \citealt{wc04}).  No standard stars were observed during the
night. For flux calibration, we analyzed $r'$ images of the X-ray
binary SAX J1808$-$3658 taken on the same night, and used a few
isolated stars in the field with calibrated photometry from the Gemini
South Telescope observations that were reported in \citet{wan+09}.

\section{Results}
\label{sec:res}

The final $r'$-band light curve of 4U~1543$-$624 is shown in
Figure~\ref{fig:lc}.  For a comparison, the light curve of 
the check star is shown in the same figure. As can be seen, periodic 
flux modulation from the LMXB is clearly visible.
The amplitude was approximately 0.1~mag: highly significant compared to
a rms deviation of 0.026~mag of the light curve of the check star.
In addition, a downward trend appears in the light curve of the binary.

Since the modulation is sinusoidal-like (see also \citealt{wc04}), we
first used a simple sinusoid to fit the light curve. A systematic
uncertainty of 0.026~mag, the standard deviation of the light curve of
the check star, was added in quadrature with the photometric
uncertainties.  The best-fit was found to have $\chi^2=197$ for 117
degrees of freedom (DoF) at period $P=18.32\pm0.09$ min.  Then a linear
function was added to the sinusoid, which gave a much improved
$\chi^2=123$ for 116 DoF.  From this we conclude that
the linear trend is highly significant.  The best-fit parameters are
$P=18.20\pm0.09$\,min, semi-amplitude $m_h=0.070\pm 0.004$\,mag, and a
linear decay of 7.0$\pm0.8\times 10^{-4}$\,mag\,min$^{-1}$.  The
best-fit function and the linear decay component are shown in
Figure~\ref{fig:lc}.

After the downward trend was removed,  the data  were
folded at the best-fit period of $P=18.20$~min. The final folded light curve is shown in
Figure~\ref{fig:fold}. As can be seen, although there are several outliers, probably due to either
intrinsic flux variations or photometry under the  
unstable observing conditions, the overall light curve shape is symmetric 
and well described
by the sinusoid. The time at the maximum of the sinusoidal fit (phase 0.5)
was MJD 54671.00545$\pm$0.00028 (Dynamical Time) at the solar system barycenter.

\section{Discussion}
\label{sec:dis}

From our time-resolved photometry, we have confirmed the presence of
the sinusoidal modulation in optical emission of 4U 1543$-$624, which
was previously reported by \citet{wc04}. The best-fit period is
18.20$\pm$0.09~min, the same as that previously obtained in 2003
August.  The semi-amplitude was 0.07~mag, slightly lower than in 2003,
but consistent within the uncertainties.  Such stability over 5 years
strongly supports the orbital origin for the modulation, indicating
that this is in fact an ultracompact binary. Unfortunately the uncertainty on
the period is large, approximately 0.5\%, preventing us from phase-connecting the two light curves; otherwise a more
accurate periodicity could be obtained (see, e.g., \citealt{wan+13}).

During our observation, we have also detected a brightness decay of
$\sim$0.1 mag over 2.3 hrs. The average magnitude was from 20.17 mag
in the beginning of our observation to 20.27 mag at the end, with an
absolute uncertainty of 0.04~mag from flux calibration. Comparing to
the average magnitude of 20.42~mag in 2003, the source was
0.15--0.25~mag brighter in 2008.  If the downward trend had continued,
the binary would have been back to 20.42~mag in another 4~hrs.
However, this decay was not seen before, and could be related to the
accretion activity in the binary system.  
Many ultracompact binaries
are known to exhibit substantial X-ray flux variations \citep{car+13}
which would have consequent changes for the flux of the irradiated companion
(although no strong X-ray variability has been reported for 4U~1543$-$624).
We thus checked the X-ray data of the binary, taken by
the All Sky Monitor onboard the \textit{Rossi X-ray Timing Explorer}
over the time of our optical observation, but did not find any evidence 
for significant X-ray variability.
We also note that it is possible to have 
a systematic 
variation in the optical brightness of an ultracompact LMXB at a time scale 
different than the binary period, without any accompanying change in 
the X-ray intensity (e.g.,\citealt{cha+01}).

The sinusoidal-like orbital modulation arises because of X-ray heating
of the inner surface of a companion star in an LMXB. The modulation amplitude 
partly depends on binary inclination $i$ and the temperature of 
the heated surface (e.g., \citealt{ak93}), and thus light-curve
fitting can provide constraints on the parameters of a binary system, such
as $i$ and source distance $d$ (see, e.g., \citealt{wan+13}). 
However, in order to fully explore the
allowed parameter space for a binary and determine the parameters with
certainty, simultaneous multiband light curves 
are needed, as the multiband flux measurements can help establish 
the temperature range for the heated surface, which would 
constrain source distance in a certain range. Given the single band of photometry
for 4U 1543$-$624, we were  only able to explore possible values
for $i$ by using a simple analytic model proposed by \citet{ak93}. 
Below we provide the estimation of the temperature for the heated
surface of the companion and the light-curve fitting in \S~\ref{subsec:lf}. 

The values of $i$ obtained from the fitting strongly depend on $d$,
which is uncertain.  \citet{wc04} estimated $d\sim 7$\,kpc by assuming
that mass transfer in the binary is driven by gravitational radiation.
Recently \citet{mad+14} have pointed out that this binary could be in
an ultra-luminous state \citep{grd09}, because of its spectral
similarities to other ultra-luminous X-ray sources. If this were the
case, the binary would be accreting near the Eddington limit and would
be at a distance of 30--40 kpc: well out in the Galactic halo.
However, we show below in \S~\ref{subsec:df} that the binary
properties are largely inconsistent with such a large distance when
considering the flux of both the companion star and the accretion disk.


\subsection{Light curve fitting}
\label{subsec:lf}

The companion star in the binary was estimated to have mass 
$M_2\simeq 0.03$~$M_{\odot}$ and radius 
$R_2\simeq 0.03$~$R_{\odot}$ \citep{wc04}.
The effective temperature of the heated side of the companion
is estimated as the following.
\citet{car+13} obtained a luminosity range (2--10 keV) of 
4.6--6.8$\times 10^{36}\ d_7^2$ erg s$^{-1}$ for this binary, 
where $d_7$ is the source distance in units of 7\,kpc, 
and \citet{mad+14} derived a slightly higher but similar luminosity (0.1--10 keV) of 
7.6$\times 10^{36}$ erg~s$^{-1}$ from their 2012 \textit{Chandra} observation.
We used the latter value as the source luminosity.
The binary separation is $D_b\simeq 1.8\times 10^{10}$~cm
for an orbital period $P= 18.2$\,min and a neutron star mass 
$M_{\rm ns}=1.4\ M_{\odot}$. The fraction $f$ of the X-ray photons
received by the companion is $f=(R_2/D_b)^2/4\simeq 0.004$.
The effective temperature of the heated side 
is $T=[f(1-\eta_{\ast})L_{\rm X}/\pi R_2^2\sigma]^{1/4}$,
where $\sigma$ is the Stefan-Boltzmann constant and 
$\eta_{\ast}$ is the albedo of the companion. 
The value for the latter is often assumed to 
be 0.5 (e.g., \citealt{ak93}), 
and thus $T\simeq $64,100$ d_7^{1/2}$~K, which we used in the following 
calculations.  Because of uncertainties on the X-ray 
flux (although it is relatively stable; \citealt{car+13}), 
the masses of the binary, and the albedo, the temperature value is
rather uncertain. We found that $\eta_{\ast}$, which
can have values of 0.6--0.8 based on the results from
modelling of emission from irradiated companions in radio pulsar
binaries or LMXBs (e.g., see \citealt{sta+01,rey+07,wan+13}),
causes the largest changes in $T$.
For example, when $\eta_{\ast}=0.8$ \citep{wan+13}, $T\simeq $51,000~K,
indicating $\sim$20\% uncertainty on $T$. 
\begin{figure}
\centering
\includegraphics[width=8cm, angle=0]{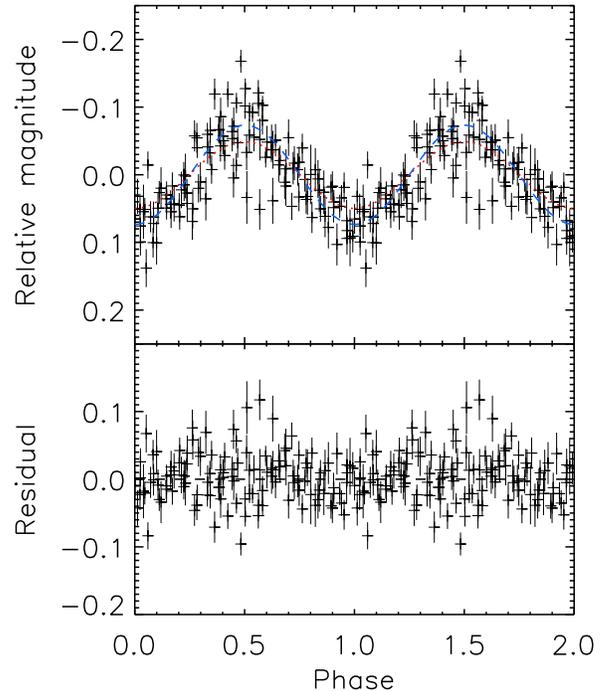}
\caption{\textit{Top panel:} The folded $r'$-band light curve of 
4U 1543$-$624 after the downward trend was removed. The (blue) dashed 
curve represents the best-fit sinusoid ($\chi^2=123$ for 116 DoF).
The (red) dotted curve is an example of
the model fit when $d=10$\,kpc and $i=82^{\circ}$ ($\chi^2=142$ for 
118 DoF).
\textit{Bottom panel:}: The residuals of the observed data points from
the best-fit sinusoid.  For clarity, two cycles are shown.}
\label{fig:fold}
\end{figure}

This is higher than the effective temperatures of most non-degenerate
stars, but hot white dwarfs can have effective temperatures as high as
200,000\,K, and at least 100 hydrogen-atmosphere white dwarfs have
effective temperatures $>$60,000\,K
(e.g., see the review by \citealt{sio11}).  Despite its very low mass
in comparison with the typical hot white dwarf, the surface
properties of the companion to 4U 1543$-$624 may otherwise be quite similar.

As the visible area of the heated side varies as a function of the
orbital phase, a modulation of $f_* [1+\sin i\sin (2\pi t/P)]$ arises,
where $t$ is time and $f_*\propto (R_2/d)^2 T^3$ is the flux from
  the irradiated side of the companion (see \citealt{ak93} for
details).  We then added a constant flux component $f_c$ to account
for the constant emission from the accretion disk, and assumed that
the temperature of the non-irradiated side of the companion was
3,000\,K, which is a value found from fitting optical light curves of
very low-mass companions in pulsar binaries (see
\citealt{sta+01,rey+07}; note that emission from this cold side is
negligible comparing to that from the accretion disk and the heated
side).  The modulation amplitude is then given by $(f_*\sin
  i)/(f_c+f_*)$ \citep{ak93}. In order to reproduce the observed
  modulation, $d$ (since $f_*\propto T^3 d^{-2} \propto d^{-1/2}$),
  $i$, and $f_c$ are key parameters to be considered.

\begin{figure}
\centering
\includegraphics[width=8cm, angle=0]{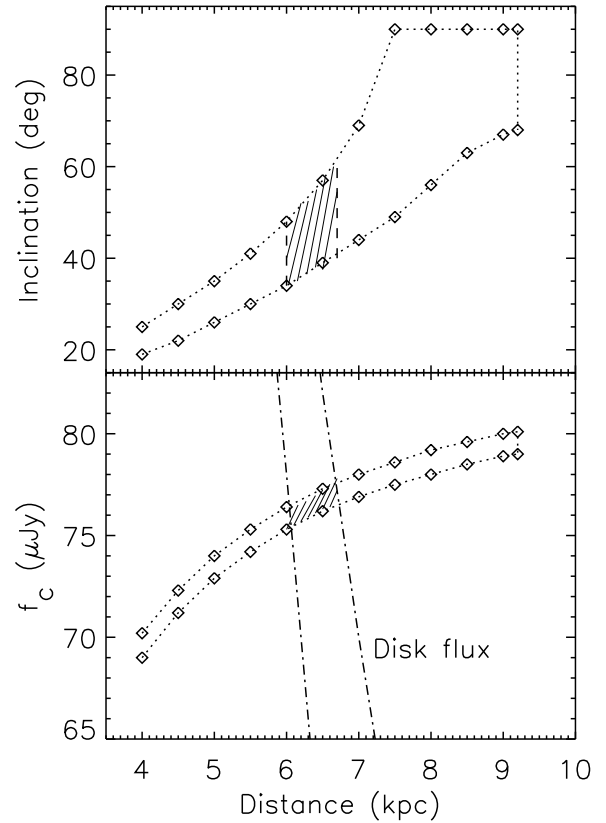}
\caption{Derived value ranges at a 90\% confidence level from light curve 
fitting (marked by diamonds and dotted curves) for inclination $i$
(top panel) and constant disk flux $f_c$ 
(bottom panel) as a function of distance $d$, where $\eta_*=0.5$ was used
(i.e., $T\simeq $64,100$ d_7^{1/2}$~K).
In the bottom panel, the disk $r'$-band $F_{\nu}$ ranges as a function of
$d$, calculated from the standard accretion disk model 
(see \S~\ref{subsec:df}), are shown by two dash-dot curves.  
The two dash-dot curves were set by using the corresponding $i$ range
at each distance in the top panel.
By requiring that $f_c$ matches 
the disk flux $F_{\nu}$ (i.e., the shaded region),
$d= 6.0$--6.7\,kpc (90\% confidence) is found, which in turn constrains
$i= 34^{\circ}$--61$^{\circ}$ (indicated by the shaded region in 
the top panel). 
\label{fig:di}}
\end{figure}

The hydrogen column density to the source is $N_{\rm H}=2.4\times
10^{21}$ cm$^{-2}$ \citep{mad+14}, which gives the extinction
$A_V\simeq 1.34$ ($A_V=N_{\rm H}/1.79\times 10^{21}$~cm$^{-2}$;
\citealt{ps95}), or $A_{r'}=1.13$ \citep{sfd98}.  We fit the
dereddened light curve of 4U~1543$-$624 to obtain constraints on $i$
and $f_c$.  The average magnitude was taken to be 20.22\,mag (obtained
from the linear decay component at the mean observing time of MJD
54672.02991): changing this slightly to account for the linear decay
would shift $f_c$ up or down fractionally by the same amount (see the
bottom panel of Figure~4).  The fitting was very degenerate with
distance, with a range of distances all having similar best-fit
$\chi^2$ values (the minimum $\chi^2\simeq 124$ for 118 DoF) up to
$d=9.2\,$kpc, beyond which no good fits could be obtained.  The
resulting $i$ and $f_c$ value ranges (90\% confidence) as a function
of $d$ are shown in Figure~\ref{fig:di}.  The best-fit model light
curve is virtually identical to the sinusoidal fit in
\S~\ref{sec:res}, and thus is not displayed in Figure~\ref{fig:fold}.
From the fitting, we found that $i$ is sensitive to distance, ranging
from 19$^{\circ}$--25$^{\circ}$ at 4\,kpc to
63$^{\circ}$--90$^{\circ}$ at 8.5\,kpc, but $f_c$ remains in the range
70--80\,$\mu$Jy. When $d > 8.5$\,kpc, the binary system is required to
be nearly edge-on. The large $\chi^2$ values indicate that $d\geq
10$\,kpc is extremely unlikely. In Figure~\ref{fig:fold}, an example
of the model light curves at $d=10$\,kpc is shown for comparison.

\subsection{Disk flux calculation}
\label{subsec:df}

The binary has a compact accretion disk and the disk flux is sensitive
to $i$ at certain distances. Here we considered the standard
geometrically thin, optically thick disk model (see \citealt{fkr02}).
The outer radius of the disk $r_o$ is approximately $r_o\simeq
1.1\times 10^{10}$\,cm, limited to the tidal radius (90\% of the
neutron star's Roche-lobe radius; \citealt{fkr02}).  The disk
temperature was determined by including both viscous heating and X-ray
irradiation (for the latter, see \citealt{vrt+90}); because of the
strong X-ray emission from the neutron star, irradiation is the
dominant heat source.  The mass accretion rate $\dot{M}$ in the disk
was estimated from $L_{\rm X}=GM_{\rm ns}\dot{M}/R_{\rm ns}$, where
the neutron star radius $R_{\rm ns}$ was assumed to be 10$^6$\,cm (at
7\,kpc distance, $\dot{M} \simeq 3.8\times 10^{16}$\,g\,s$^{-1}$).
The disk flux $F_{\nu}$ at frequency $\nu$ is then given as \citep{fkr02}
\[
F_{\nu}= \frac{4\pi h\nu^3\cos i}{c^2 d^2}\int^{r_o}_{r_i}\frac{rdr}{e^{h\nu/k_BT_d}-1}\ ,
\]
where $h$ is Planck's constant, $k_B$ Boltzmann's constant, and
  $c$ the speed of light. The disk temperature $T_d$ is a function of
the disk radius $r$, $T_d \propto \dot{M}^{2/7} r^{-3/7}$
  \citep{vrt+90}, and $r_i$ is the inner disk radius (we set
$r_i=10^7$\,cm in our calculation, but note that the disk flux is not
sensitive to $r_i$ as long as $r_i\ll r_o$).  We calculated the disk
flux at the central wavelength (6300\,\AA) of the $r'$ band over the
same range of distance as considered in Figure~\ref{fig:di}, where we
required that $i(d)$ matched the constraints from \S~\ref{subsec:lf}.
The possible range of $F_\nu$ is shown by the two
  dash-dot curves in the bottom panel of Figure~\ref{fig:di}.
Requiring the disk flux $F_{\nu}$ to be consistent with $f_c$ from \S~\ref{subsec:lf}, we find $d= 6.0$--6.7\,kpc and $i=
34^{\circ}$--61$^{\circ}$ (90\% confidence).

We note that the $i$ value range is
lower than that of $65^{\circ}$--82$^{\circ}$ obtained by
\citet{mad+14} from their X-ray spectral fitting, but are consistent
with the non-detection of any eclipses or dips at X-ray energies
(which generally suggest $i<60^{\circ}$; \citealt{fkr02}).  
However our results were calculated from simple analytic models
for X-ray binaries and depend on parameters such as $\eta_{\ast}$ that
were not robustly examined here. For example, if $\eta_{\ast} = 0.8$
(i.e., $T\simeq 51,000 d^{1/2}_7$\,K) is assumed, the above calculations 
will result in $d\sim 5.2$--6.0\,kpc and $i\sim 35$--87$^{\circ}$.
In any
case, our calculations clearly show that to produce the optical
modulation seen in 4U~1543$-$624 at optical wavelengths, it is likely
that $d\leq 9$\,kpc, not supporting the suggestion that this binary is
an ultra-luminous X-ray source. Instead, our fitting suggests $d\sim
6.0$--6.7\,kpc.  Following similar calculations presented here, future
multiband observations will certainly help our study of this
ultracompact binary by determining its distance and inclination.



\begin{acknowledgements}
We thank the anonymous referee for constructive suggestions.
This research was supported by the National Natural Science Foundation of
China (11373055) and the Strategic Priority Research Program
``The Emergence of Cosmological Structures" of the Chinese Academy
of Sciences (Grant No. XDB09000000). 
A.T. acknowledges support from Chinese Academy of Sciences visiting 
Fellowship for Researchers from Developing Countries. 
\end{acknowledgements}

\bibliographystyle{apj}

\end{document}